\documentclass[]{aastex631}
\usepackage{multirow}

\newcommand{\Add}[1]{#1}

\shorttitle{AASTeX v6.3.1 Sample article}
\shortauthors{Yutani et al.}
\graphicspath{{./}{figs/}}

\begin{document}

\title{Origin and evolution of dust-obscured galaxies in galaxy mergers}

\author{Naomichi Yutani}
\affiliation{Kagoshima University, Graduate School of Science and Engineering, Kagoshima 890-0065, Japan}
\email{yutaninm@gmail.com}

\author[0000-0002-3531-7863]{Yoshiki Toba}
\affiliation{National Astronomical Observatory of Japan, 2-21-1 Osawa, Mitaka, Tokyo 181-8588, Japan}
\affiliation{Department of Astronomy, Kyoto University, Kitashirakawa-Oiwake-cho, Sakyo-ku, Kyoto 606-8502, Japan}
\affiliation{Academia Sinica Institute of Astronomy and Astrophysics, 11F of Astronomy-Mathematics Building, AS/NTU, No.1,\\ Section 4,Roosevelt Road, Taipei 10617, Taiwan}
\affiliation{Research Center for Space and Cosmic Evolution, Ehime University, 2-5 Bunkyo-cho, Matsuyama, Ehime 790-8577, Japan}

\author[0000-0002-9850-6290]{Shunsuke Baba}
\affiliation{Kagoshima University, Graduate School of Science and Engineering, Kagoshima 890-0065, Japan}

\author[0000-0002-8779-8486]{Keiichi Wada}
\affiliation{Kagoshima University, Graduate School of Science and Engineering, Kagoshima 890-0065, Japan}
\affiliation{Research Center for Space and Cosmic Evolution, Ehime University, 2-5 Bunkyo-cho, Matsuyama, Ehime 790-8577, Japan}
\affiliation{Hokkaido University, Faculty of Science, Sapporo 060-0810, Japan}
\begin{abstract}
	Dust Obscured Galaxies (DOGs), which are observationally characterized as faint in the optical and bright in the infrared, are the final stage of galaxy mergers and are essential objects in the evolution of galaxies and active galactic nuclei (AGNs). 
	However, the relationship between torus-scale gas dynamics around AGNs and DOGs lifetime remain unclear.
	We obtained evolution of the spectral energy distributions (SEDs) of a galaxy merger system with AGN feedback, from post-processed pseudo-observations based on an N-body/Smoothed Particle Hydrodynamics (SPH) simulation.
	We focused on a late stage merger of two identical galaxies with a supermassive black hole (SMBH) of 10$^8$ M$_\odot$. 
	We found that the infrared luminosity of the system reaches ultra- and hyper-luminous infrared galaxy classes (10$^{12}$ and 10$^{13}$ L$_\odot$, respectively).
	The DOGs phase corresponds to a state in which the AGNs are buried in dense gas and dust, with the infrared luminosity exceeding 3.3 $\times$ 10$^{12}$ L$_\odot$.
	We also identified the sub-categories of DOGs, namely bump and power-law DOGs from the SEDs and their evolution.
	The bump DOGs tend to evolve to power-law DOGs on several Myrs.
	We found that contribution from the hot dust around the nucleus in the infrared radiation is essential for identifying the system as a power-law DOG; the gas and dust distribute non-spherically around the nucleus, therefore, the observed properties of DOGs depend on the viewing angle.
	In our model, the lifetime of merger-driven DOGs is less than 4 Myrs, suggesting that the observed DOGs phase is a brief aspect of galaxy mergers.
\end{abstract}



\section{Introduction} \label{sec:intro}
Galactic collisions have received much attention as a possible explanation for the formation of supermassive black holes (SMBHs) and their parent galaxy relationships.
For example, \citet{hopkins2008} presented a scenario of ultra-luminous and luminous infrared galaxy (U/LIRG)\footnote{LIRGs are galaxies with total infrared emissions brighter than $10^{11}\rm{\ L_{\odot}}$ and less than $10^{12}\rm{\ L_{\odot}}$ and ULIRGs with total infrared emissions of $10^{12}\rm{\ L_{\odot}}$ or larger.} formation by major mergers.
In gas-rich mergers, large amounts of gas and dust fall into the center of the galaxies because of galaxy-galaxy interactions, triggering nuclear starbursts and the active galactic nuclei (AGNs).
AGNs will expectedly be buried in a dense interstellar medium (ISM) for at least a certain period during this process \citep{hickox2018}.
In fact, \citet{ricci2017} claimed that galaxies in the late phase of mergers have a higher fraction of Compton-thick AGNs with hydrogen column density ($N_{\rm H}\,\geq\,2\,\times\,10^{24}\,{\rm cm}^{-2}$) than those that are isolated or in the early merging phase \citep[see also][]{yamada2021}. 
The mass accretion rate to the nuclei can be enhanced during the buried phase \citep{kawaguchi2020}, which is also essential for understanding the formation of SMBHs during galaxy formation.\par

Dust-obscured galaxies (DOGs) could be representative galaxies in the late stages of galaxy mergers, and they could evolve into quasars or red elliptical galaxies \citep{dey2008}. 
This scenario is consistent with the fact that the population density of DOGs and quasars, as well as the cosmic star formation rate (SFR) commonly peak around the redshift z $\sim$ 2 \citep{hopkins2004,hopkins2007}.
DOGs are characterized as being faint in the optical and bright in the infrared; throughout this paper, we use the criteria of the color between R-band to Spitzer 24 $\rm \mu m$ band R$-[24]\,>\,14.0$ in Vega magnitude, which is equivalent to $F_{24\mu m}/F_{\rm R}\geq982$, and $Spitzer$ 24 $\rm{\mu m}$ band flux $F_{24\rm{\mu m}}\,\geq\,0.3\rm{\,mJy}$ \citep{dey2008}.\par

DOGs can be divided into two sub-categories, namely, ``bump DOGs" and ``power-law DOGs" (hereafter PL DOGs) \citep{dey2008}.
The spectrum of bump DOGs is characterized by a bump in the spectral energy distribution (SED) at 1.6 $\rm \mu$m and polycyclic aromatic hydrocarbon (PAH) emissions.
Alternately, PL DOGs are characterized by a power-law spectrum in the optical and infrared wavelengths.
Although their origin is still unclear, it is often considered that active star formation contributes to the SEDs of bump DOGs \citep{dey2008,melbourne2012}.
It has also been reported that the population ratio of PL DOGs to bump DOGs correlates to $F_{24\rm{\mu m}}$ \citep{dey2008,melbourne2012,toba2015}.
In addition, PL DOGs tend to have weaker PAH emissions and a higher $F_{24\rm{\mu m}}$ than bump DOGs.
These weak PAH emissions are usually regarded as a sign of AGN activity \citep{houck2005,yan2007}, and thus explains why PL DOGs are often considered as AGN dominated objects.
However, this interpretation was not statistically confirmed by X-ray surveys of DOGs.
The structure in the central kpc region has not elucidated due to DOGs being mainly found at high redshift (z $>$ 1).
Stated differently, no observed evidence confirms how the AGNs, if present in the core of DOGs, are obscured by gas and dust, whose distribution is generally not spherically symmetric.
Ideally, the color and \Add{infrared (IR)} flux should be affected by the geometry and distribution of the gas in the central 100 pc.
The difference between PL DOGs and bump DOGs in terms of their structures is also an open question.
Moreover, there is no theoretical understanding of the relationship between AGN activity and PL DOGs.\par

Numerical modelling of the galaxy-galaxy merger helps to study the origin of DOGs.
\Add{For example, \citet{narayanan2010} performed galaxy-galaxy merger simulations at redshift 2.}
\Add{They found that the mid-IR SED of the system evolves from bump-type to power-law type during the final coalescence stage.}
However, the spatial resolution from previous simulations of gas-rich mergers are not necessarily fine enough to resolve the dusty torus in which SMBHs are deeply buried.
\citet{narayanan2010} used a 100 pc resolution; therefore, it was necessary to introduce a subgrid model to represent the gas dynamics and structures around the SMBH \citep[see also][]{crain2015}.
\citet{blecha2018} studied the evolution of gas-rich merger systems using the N-body/Smoothed Particle Hydrodynamics (SPH) code \texttt{GAGEDT-3} with a spatial resolution of 23--48 pc and a mass resolution of $2.8\times10^5$ ${\rm M_{\odot}}$.
More recently, \citet{yang2021} performed similar simulations for less massive galaxies with smaller BHs ($1.5-3\times10^6$ ${\rm M_{\odot}}$) using \texttt{GADGET-2} with spatial resolutions and masses of 20 pc and $4.6\times10^3$ ${\rm M_{\odot}}$, respectively.
The geometry and internal distribution of the gas and dust in $r\,<\,100 $ pc are crucial for multiwavelength observational properties of AGNs \citep{schartmann2014,wada2016,izumi2018}.
For example, molecular gas around AGNs has recently been resolved by ALMA \citep{garcia2016,izumi2018,combes2019}, which showed that the size of the molecular tori is typically less than 30 pc, and they are characterized by complicated kinematic structures \citep{imanishi2018,imanishi2021}.
Moreover, the mass accretion rate to the SMBH can be affected by AGN feedback at r $\sim$ several tens of pc \citep{kawaguchi2020}.
Therefore, gas dynamics resolution in the torus-scale (i.e., several tens of parsecs ) is required to understand the origin of DOGs.\par

In this study, we focused on the final phase of galaxy mergers to reveal how the shielding of SMBHs evolve with time.
We modeled the interaction of approximately 1 kpc cores in two galactic centers (see Section \ref{sec:models}), assuming that the mass accretion toward the galactic nuclei is dominated by dense gas in the galactic central region.
This ensures better spatial (4 pc) and SPH mass (1000 M$_\odot$) resolutions than previous gas-rich major merger simulations \citep[for example][]{narayanan2010,blecha2018,yang2021}.\footnote{In the most recent cosmological simulations of quasar fueling, \citet{angles2021} attained a sub-pc spatial resolution in a $10^{12} {\rm M_{\odot}}$ halo using a Lagrangial refinement technique. Their model is intended to model the fueling of a single SMBH without AGN feedback and does not treat galactic collisions.}
The formation and evolution of the merger system were studied by post-processed pseudo-observations using the radiation transfer simulation code \texttt{RADMC-3D} \citep{dullemond2012}.
\Add{Note that we do not fully resolve the gas dynamics on the accretion disk scale in the late stages of galaxy mergers. However, with the spatial resolution in our simulations, we can resolve three-dimensional structures of the gas on the dust-torus scale, which is essential for the obscuring properties of the AGNs. Then we can discuss, for example, how the properties of DOGs depend on the line-of-sight directions. We can also investigate the evolution of SEDs during the mergers with finner time resolution than in previous simulations.}\par

The remainder of this paper is organized as follows.
Section \ref{sec:models&methods} describes the simulation methods and the models.
In Section \ref{sec:macc}, we investigate the time evolution of gas density and mass accretion rates.
In Sections \ref{sec:sed} and \ref{sec:agn-activity}, we present pseudo-observation results.
The evolution of the infrared luminosity and color calculated based on SED and the column density during the DOGs phase and the relationship between the bump and PL DOGs are discussed in Section \ref{sec:discuss}.\par

Unless otherwise noted, all magnitudes in this study refer to the Vega system.
In addition, a flat universe of
$H_0\,=\,70\,\rm{km s^{-1} \rm{Mpc^{-1}}},\Omega_{M}\,=\,0.3,$ and 
$ {\Omega}_{\lambda}\,=\,0.7$ was assumed.\par

\section{Models and methods} \label{sec:models&methods}
We simulated the interaction of galactic nuclei with supermassive black holes (SMBHs) in the final phase of a galaxy merger using N-body/SPH simulation code \texttt{ASURA} \citep{saitoh2008,saitoh2009,saitoh2013}.
The results obtained from \texttt{ASURA} were used as inputs for \texttt{RADMC-3D} \citep{dullemond2012} to calculate the evolution of SEDs.

\begin{deluxetable*}{cccccccccccccccc}
	\tablenum{1}
	\tablecaption{Parameters of the pre-merger system\label{tab:model_parameter}}
	\tablewidth{0pt}
	\tablehead{
		\multicolumn{2}{c}{SMBH} && \multicolumn{4}{c}{Stellar system} && \multicolumn{4}{c}{Gas disk} & \multirow{2}{*}{$\epsilon$$^{\rm k}$} & \multirow{2}{*}{${\eta_{\rm AGN}}^{\rm l}$}\\
		\cline{1-2}\cline{4-7}\cline{9-12}
		${M_{\rm BH}}^{\rm a}$ & ${r_{acc}}^{\rm b}$ && ${M_{star}}^{\rm c}$ & ${R_{c}}^{\rm d}$ & ${N_{star}}^{\rm e}$ & ${\Delta M_{star}}^{\rm f}$ && ${M_{\rm gas}}^{\rm g}$ & ${R_{\rm gas}}^{\rm h}$ & ${N_{gas}}^{\rm i}$ & ${\Delta M_{\rm gas}}^{\rm j}$ & & &
	}
	\startdata
	$10^8 {\rm M_{\odot}}$ &  6.4 pc && $2\times10^{10} {\rm M_{\odot}}$ & 800 pc & $2\times10^6$ & $10^4 M_{\odot}$ && $4\times10^9 {\rm M_{\odot}}$ & 800 kpc & $4\times10^6$ & $10^3 {\rm M_{\odot}}$ & 4.0 pc & 0.2\%\\
	\enddata
	\tablecomments{
		(a) Mass and (b) accretion radius of a sink particle. 
		(c) Total mass and (d) core radius of Plummer sphere. 
		(e) Number of particles in the spherical stellar system. 
		(f) Mass resolution of star particles in a spherical stellar system. 
		(g) Total mass and (h) outer edge radius of uniform-density gas disk. 
		(i) Total number and (j) mass resolution of the gas disk particles. 
		(k) Gravitational softening. 
		(l) AGN feedback efficiency.}
\end{deluxetable*}

\subsection{Merger simulations}\label{sec:asura}
\subsubsection{N-body/SPH simulations by ASURA}
The dynamics of the interstellar medium in merger system are calculated using smoothed pseudo-density SPH \citep[SPSPH;][]{yamamoto2015}.
Gravitational forces were computed using a parallel tree algorithm.
The tolerence parameter was set to 0.5.\par

Star formation and supernova explosions were implemented based on \citet{saitoh2008}.
The four conditions under which SPH particles are converted into star particles are 
(1) Hydrogen number density $n_{\rm H}\,>\,400\rm{\,cm^{-3}}$; 
(2) Gas temperature $T_{gas}\,<\,100\rm{\,K}$; 
(3) $\nabla\cdot {\mathbf{v}}_{SPH}\,<\,0$, where ${\mathbf{v}}_{SPH}$ is the velocity of the SPH particle.
(4) $\Delta Q\,<\,0$, where $\Delta Q$ is the thermal energy received by an SPH particle in a time step.
SPH particles satisfying these conditions are converted to star particles according to the local Schmidt's law \citep{scmidt1959}.
The star formation rate (SFR) is determined as $C_* \rho_{gas}/t_{ff}$, where $C_*$ is a dimensionless star-formation efficiency parameter set to 0.033 herein following \cite{saitoh2008}, $\rho_{gas}$ is the local gas density, and $t_{ff}$ is the free fall time.
The initial mass function of the stellar particle was chosen as in \citet{salpeter1955}. 
The supernova explosion was introduced based on \citet{okamoto2008}, and the stellar particles underwent a probabilistic Type II supernova explosion. 
We assume that a single supernova explosion provides $10^{51}\rm{\,erg}$ as thermal energy to the surrounding SPH particles weighted by the spline function.
Optically thin radiative cooling was assumed for the gas at $10\rm{\,K}\,$--$\,10^8\rm{\,K}$ \citep{wada2009} with solar metallicity.
Far-ultraviolet radiation and photoelectric heating were also considered.\par

\subsubsection{Merger models}\label{sec:models}
\begin{figure}[ht!]
	\begin{center}
		\includegraphics[width=8cm, angle=270, bb=0 0 158mm 214mm]{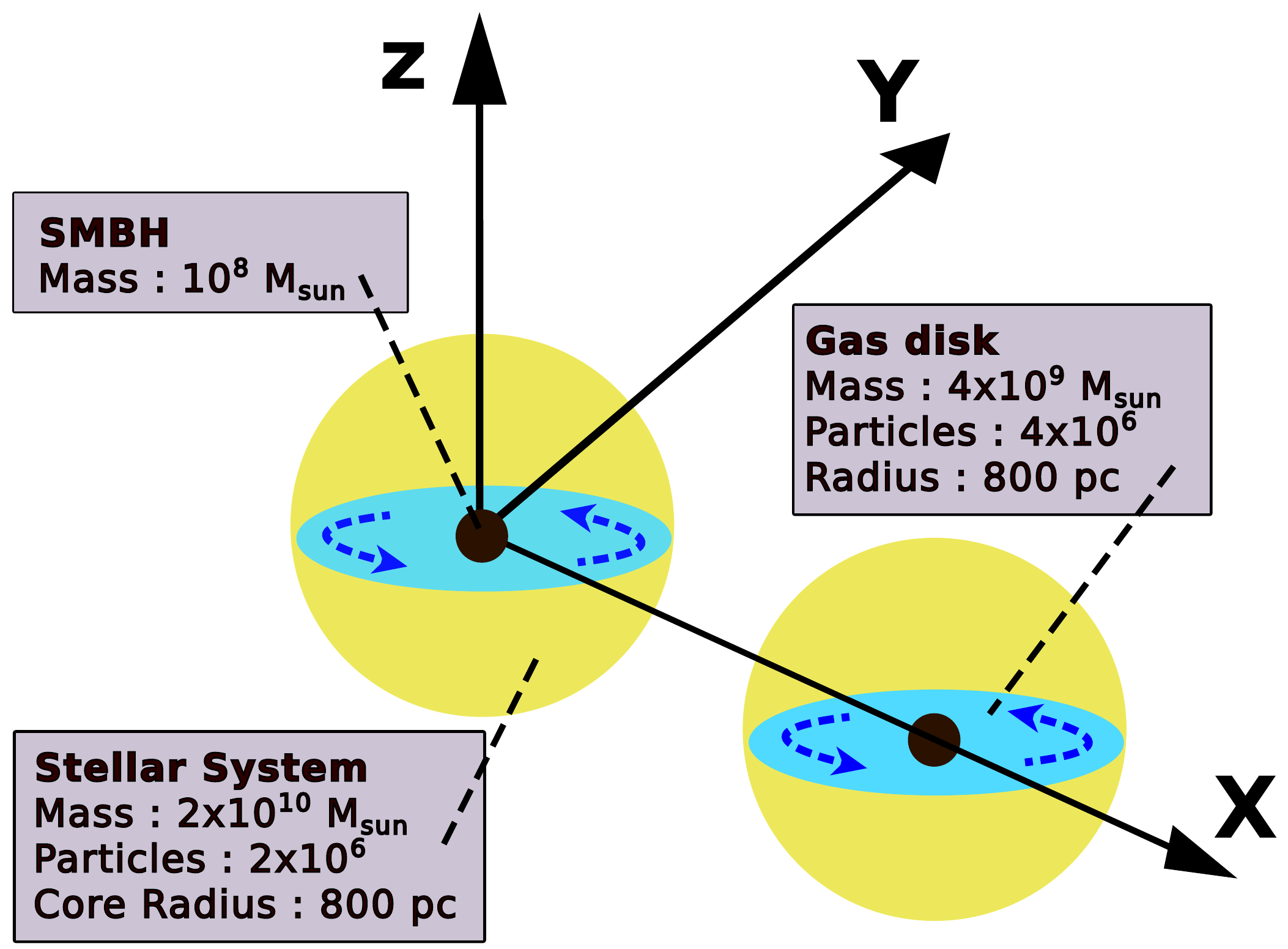}
	\end{center}
	\caption{Initial setup of the merger system. The two identical systems were merged.
		Each has a supermassive black hole (SMBH) of $10^8\,\rm{M_{\odot}}$, a rotationally supported gas disk with a mass of 4 $\times$ $10^9\, \rm{M_{\odot}}$ and angular momentum in the z-axis direction, and a spherical stellar system of $2\,\times\,10^{10}\rm{\ M_{\odot}}$. \label{fig:model}}
\end{figure}

Here, we focus on the final phase of a galaxy-galaxy merger to study how the two BHs interact with the gas in the central kpc region. 
We did not consider galactic disks or dark matter halos.
The model is based on \citet{kawaguchi2020}, who studied the interactions between SMBHs and ISM in the central subkiloparsec region using the N-body/SPH code \texttt{ASURA}.
As an initial condition, we prepared two identical systems, as schematically represented in Figure \ref{fig:model}.
Each system consists of three components: a SMBH ($M_{\rm BH}\,=\,10^8\rm{\ M_{\odot}}$), spherical stellar system ($M_{star}\,=\,2\,\times\,10^{10}\rm{\ M_{\odot}}$), and axisymmetric rotating gas disk ($M_{\rm gas}$ = 4 $\times\,10^9\rm{\ M_{\odot}}$) (see Figure \ref{fig:model}).

The initial stellar system has a Plummer mass distribution, and the initial gas disk has a uniform density and a radius of 800 pc in rotational equilibrium \citep[for example][]{sorai2002,herrero2012}.
We posit that mass accretion toward the galactic nucleus is dominated by dense gas and stars in the galactic central kpc region.
The stellar system and gas disk are represented by $2\times10^6$ N-body and  4 $\times\ 10^6$ SPH particles, respectively.
We represent a stellar sphere assuming a Plummer model (i.e., the gravitational potential $\Phi(r) = -GM_{star}/\sqrt{r^2+R_{c}^2}$ ) with $M_{star}=2\times10^{10}\rm{\ M_{\odot}}$ and core radius $R_{c} = 800$ pc.\par

The initial positions and velocities for each system are listed in Table \ref{tab:pasi_and_vel}.
The gas disks of the two identical systems were coplanar and their spins were prograde.
Here, we define a cartesian coordinate such that the origin is the center of one system (System 1), the x-axis is the direction to the other system (System 2), and the z-axis is parallel to the disk spins.
System 2 is initially at ($x$, $y$, $z$) = (3.5, 0, 0) kpc and has ($v_x$, $v_y$, $v_z$) = (0, 100, 0) pc Myr$^{-1}$ (Table \ref{tab:pasi_and_vel}).
Therefore, the entire merging system's orbit is coplanar with the gas disks (x-y plane), and its rotational axis is parallel and prograde with the gas disk spins (z-axis).
\citet{kawaguchi2020} found that there is no significant difference between prograde and retrograde collisions in terms of the mass accretion toward SMBHs after studying collisions between the galactic cores with various orbits and spin vectors using a similar numerical method as in this paper.
Therefore, we focused on prograde collisions.
\par

\begin{deluxetable*}{ccc}
	\tablenum{2}
	\tablecaption{Relative velocity and position\label{tab:pasi_and_vel}}
	\tablewidth{0pt}
	\tablehead{
        & Position [kpc] & Velocity [pc/Myr]
	}
	\startdata
	System1 & (0,0,0) & (0,0,0) \\
	System2 & (3.5,0,0) & (0,100,0) \\
	\enddata
\end{deluxetable*}

\subsubsection{ Mass accretion and AGN feedback }
An SPH particle is assumed to have accreted to a BH when the SPH particle enters the accretion radius ($r_{acc}$) from the BH particle, and the following two conditions are satisfied: 
1) The kinetic energy of the SPH particle is smaller than the gravitational energy and 
2) The angular momentum of the SPH particle is smaller than $J_{acc}\,=\,r_{acc}\sqrt{GM_{\rm BH}/(r_{acc}^2\,+\,\epsilon^2)^{1/2}}$, where $\epsilon$ is gravitational softening.
In this study, we assumed $\epsilon$ = 4.0 pc and $r_{acc}$ = 6.4 pc.
\Add{\textit{The accretion radius should be larger than softening radius to well resolve accretion process.}
However, if $r_{acc}$ is larger than the typical torus scale ($\sim$ 10 -- 20 pc), the accretion rate toward the sink particle would be overestimated.
We then assume $r_{acc}$ = 6.4 pc to resolve the gas dynamics near the gravitational softening radius.}\par

The \Add{isotropic} AGN feedback provides the thermal energy ($\Delta E\,=\,\eta_{AGN}\dot{M}_{\rm gas}c^2$) weighted by the spline function to 96 -- 272 SPH particles in the vicinity of each BH particle.
\Add{The AGN feedback radius is defined as the number of SPH particles within the radius that satisfies a range (96-272, the average is 145)\footnote{\Add{To seek the exact number of SPH particles every timestep computationally costs, especially for dense regions near sink particles.}}.}
\Add{The $\dot{M}_{\rm gas}$ is the gas mass accretion rate of a BH particle per step at $r_{acc}$.}
\Add{The} $\eta_{AGN}$ is a free parameter representing the energy-loading efficiency, that is, a fraction of the heat energy received by the SPH particles in an arbitrary radius around the BH particles.
Note that the AGN feedback implemented here does not exactly model the radiative and mechanical feedback from the AGNs to the circumnuclear gas in real galaxies.
In other words, the parameter is not determined from the first principle.
This parameter phenomenologically expresses the feedback process in a finite spatial resolution (i.e., 4 pc in this study).
The relevant feedback fraction depends on the numerical methods, spatial resolution, and equation of state of the ISM; therefore, it is determined heuristically \citep[see for example, ][]{hopkins2016}.
\citet{kawaguchi2020} suggested in their N-body/SPH simulations of mergers that when $\eta_{AGN}\,=\,2.0$\%, the gas around the BH particles is blown away by the energy injection associated with the mass accretion; as a result, the mass accretion to the BH particles cannot be maintained.
They found that mass accretion and AGN feedback could coexist for $\eta_{AGN}\,=\,0.2$\%.
We also used this number in this study.
The physical scale in which thermal energy is supplied is usually several tens of parsecs, depending on the gas density around the BH particles.
However, in high-density regions, radiative cooling exceeds AGN feedback heating.
\Add{In our simulation, the cooling of SPH particles that receive AGN feedback energy was set to be inactive for 10 time steps to ensure the hot gas produced by the AGN-feedback to interact with its surroundings.}
\Add{This short period (10 steps = 10--10$^3$ yr) for which the cooling is stopped can be justified because it is shorter than the sound crossing time of each SPH particle\footnote{\Add{We confirmed that the AGN luminosity is not affected when the cooling is stopped for 100 steps.}}.}

\subsection{ Radiative transfer simulations }\label{sec:radmc}
Snapshots obtained from the merger simulations described in Section \ref{sec:asura} were used to calculate the SED evolution using \texttt{RADMC-3D}, which is a Monte Carlo radiation transport code \citep{dullemond2012}.\par

In this study, we considered two types of radiation sources: AGNs associated with the two SMBHs, and star clusters consisting of newly formed young stars and bulge stars.
The spectrum of each AGN depends on the bolometric luminosity $L_{bol}\,=\,\varepsilon\dot{M}_{\rm gas}c^2$, where: The energy conversion efficiency $\varepsilon = 0.1$, and $\dot{M}_{\rm gas}$ is the \Add{gas} mass accretion rate of the SMBH averaged over 0.1 Myr.\par

The spectrum of each AGN \Add{was} obtained from the templates contained in the \texttt{Cloudy} code \citep{ferland1998}.
\Add{In the code, we set four parameters to define a typical AGN SED consisting of multiple continuum components as follows.}
\Add{Firstly, the temperature of the Big Blue bump set to 1.5 $\times$ 10$^5$ K.}
\Add{Secondly, the X-ray to UV ratio $\alpha_{ox}$ was assumed to be -1.4, which is a typical value in AGN \citep{zamorani1981,lusso2010}.}
\Add{In addition, we assumed that $\alpha_{uv}$, which determines the low-energy slope of the Big Blue bump, and $\alpha_{x}$, which determines the slope of the power-law X-ray component, were -0.5 \citep{francis1993,natali1998} and -1.0 \citep{mateos2010,waddell2020}, respectively.}
For stellar clusters, we \Add{assumed} Single Stellar Population (SSP) models; their spectra \Add{was} obtained using \texttt{P\'egase.3} \citep{fioc2019}.
The initial mass function of an SSP \Add{set} \citet{salpeter1955}, and the spectrum of the SSP depends on the formation time.\par

We assumed that the spatial distributions of gas and dust were the same as the gas-to-dust mass ratio of 50 \citep{toba2017a}.
The size distribution of the dust grains was obtained from \citet{weingartner2001} ($R_v\,=\,3.1$).
This model includes silicate, graphite, and neutral and ionized PAHs.
Opacity curves were obtained from \citet{laor1993}.\footnote{data available at ftp://ftp.astro.princeton.edu/draine/dust/diel/}
The mass ratio of graphite to silicate was assumed as $M_{\rm gra}/M_{\rm sil}$ = 1.0.\par

In \texttt{RADMC-3D}, we must specify three types of photon numbers ($N_{therm}$, $N_{scat}$, and $\,N_{spec}$), where $N_{therm}$ is used to determine the temperature distribution of the dust, $N_{scat}$ is used to calculate the scattered light when calculating the intensity distribution, and $N_{spec}$ is used to obtain the flux density. 
In this study, we assume that $N_{therm}=10^6$, $N_{scat}=10^6$, and $N_{spec}=10^5$.
The starting value of the random seed for the Monte Carlo simulation was used with default (-17933201).
We have confirmed that the SEDs do not change significantly even if all three photon numbers are increased by a factor of five.\par

\section{ Results }\label{sec:results}
\subsection{Mass accretion to BHs}\label{sec:macc}
\begin{figure}[ht!]
	\begin{center}
		\includegraphics[width=18cm,bb=0 0 19cm 12cm]{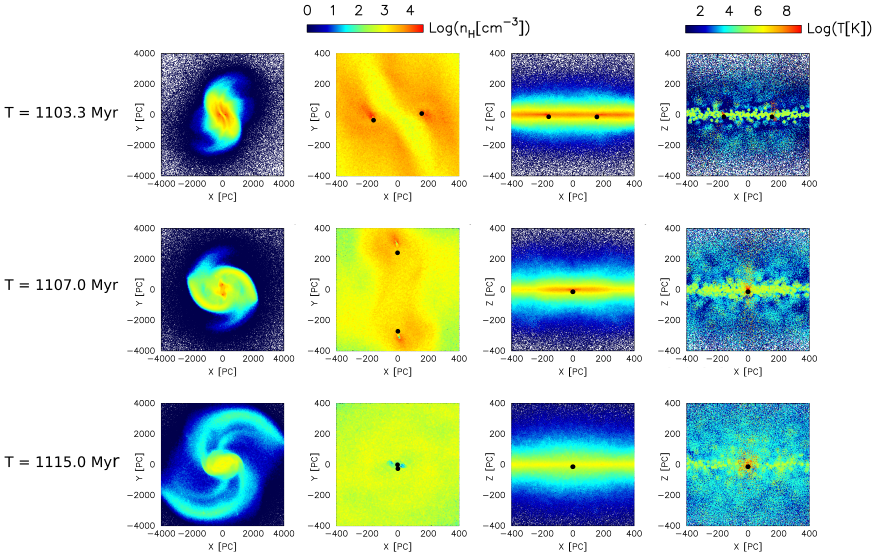}
	\end{center}
	\caption{Gas density distribution ($n_{\rm H} [\rm{cm^{-3}}]$) 
		on the first left to third columns and gas temperature ($\rm T[K]$) on the right end column.
		From top to bottom, $t =$ 1103.3, 1107.0, and 1115.0 Myr.
		Note that, here, the coordinates are shifted parallel to those in Figure \ref{fig:model} such that the origin is the center of gravity of the two BHs.
		Black circles denote the positions of the two BHs.\label{fig:gas}
	}
\end{figure}

Figure \ref{fig:gas} shows the gas density and temperature distributions from top to bottom t = 1103.3, 1107.0, and 1115.0 Myr.
The first left column represents gas density projected onto the x-y plane in 8 $\times$ 8 kpc$^2$, the evolution of the tidal tail due to collisions between the galactic cores.
The second and third left column also represents gas density projected on the the x-y plane and x-z plane in 800 $\times$ 800 pc$^2$, respectively.
From 1103.3 to 1115.0 Myr, the decrease in density and increase in scale height of the gas disk can be observed.
The right end column is gas temperature projected onto x-z plane in 800 $\times$ 800 pc$^2$.
Local heating by SN feedback can be confirmed.
Although isotropic AGN feedback is implemented, we can observe a mass outflow driven by the AGN feedback in the z-direction from the gas temperature distribution at t = 1107 and 1115 Myr.\par

\begin{figure}[ht!]
	\begin{center}
		\includegraphics[width=8cm, bb=0 0 190mm 381mm]{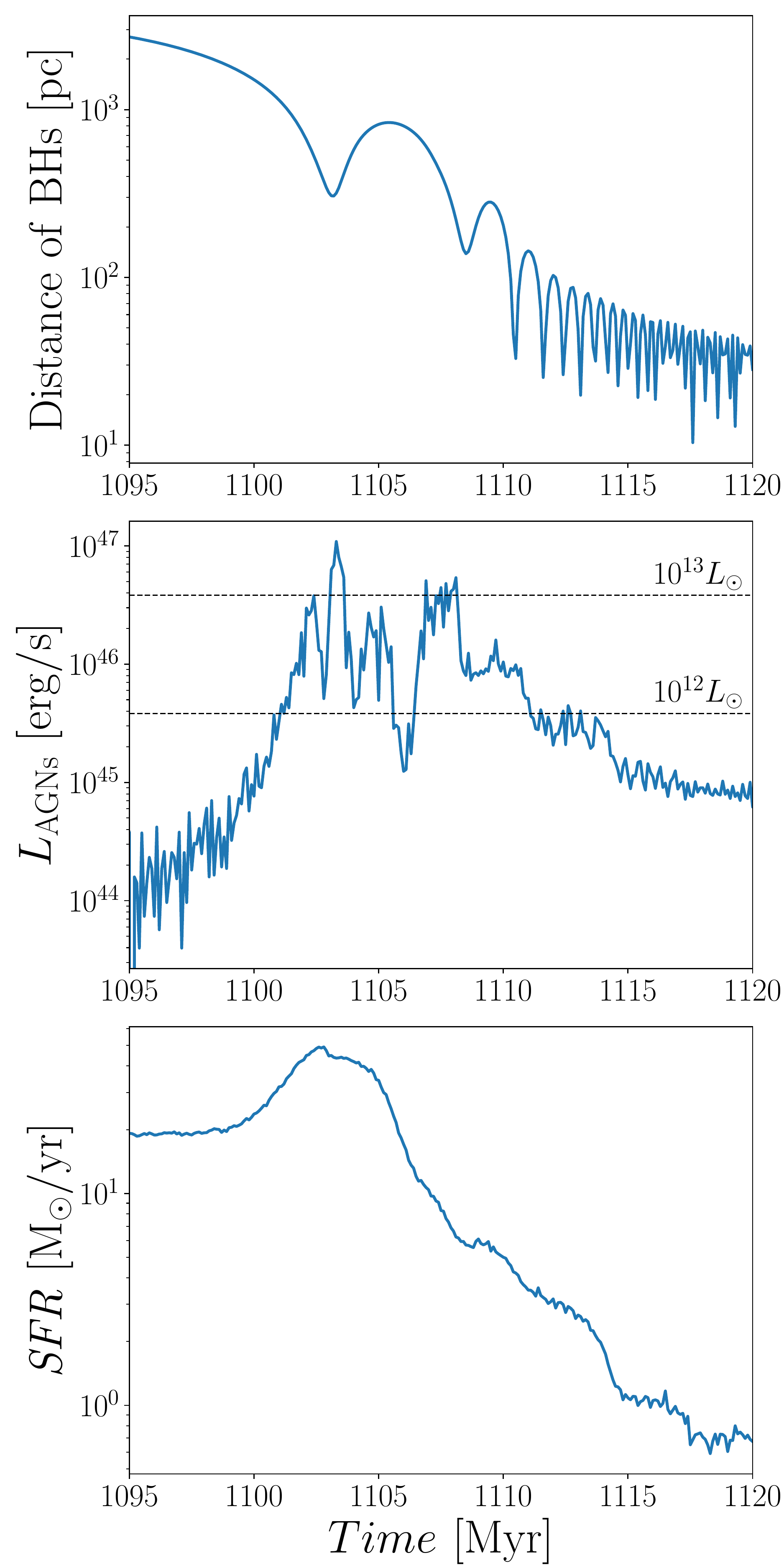}
	\end{center}
	\caption{The time evolution of the distance between the two BHs (top panel), total bolometric luminosity $L_{AGNs}=0.1\times\dot{M}_{\rm gas}c^2$ of the two AGNs (middle panel), Star formation rate (SFR) (bottom panel).\label{fig:bhdis&macc}}
\end{figure}

Figure \ref{fig:bhdis&macc} shows the BH-BH separation, total AGN luminosity, and star formation rate (SFR) of the system, as a function of time.
As shown in the top panel of Figure \ref{fig:bhdis&macc}, the distance between the two BHs gradually approaches each other while oscillating.
The total bolometric luminosity $L_{\rm AGNs}=0.1\times\dot{M}_{\rm gas}c^2$ of the two AGNs reached a maximum of 1.09 $\times$ 10$^{47}$ erg$\rm\,s^{-1}$ at t = 1103.3 Myr (middle panel of Figure \ref{fig:bhdis&macc}).
In this active phase, high-density regions ($n_{\rm H}>10^{3.5}\ \rm{cm^{-3}}$) were formed by the interactions between the two systems (see Figure\ref{fig:gas}).
At 1102.8 Myr, the SFR reaches a maximum of 49 ${\rm M_\odot\,yr^{-1}}$ (bottom panel of Figure \ref{fig:bhdis&macc}).
We found that $L_{\rm AGNs}$ exceeds 3.828 $\times$ 10$^{46}$ erg s$^{-1}$ (i.e., $10^{13}$ $L_{\rm\odot}$) at approximately 1103 Myr.
After $t=$ 1110 Myr, $L_{\rm AGNs}$ decreased monotonically.
After 1115 Myr, the two systems became relaxed, and the AGN luminosity decreased to less than 10$^{46}$ erg s$^{-1}$.
The gas in the disks is lost because of star formation, thermal AGN feedback, and angular momentum transportation in the active phase, and the average density of the disks is less than $10^3\ \rm{cm^{-3}}$ in the relaxed phase.
In the following analysis, because we are interested in the formation and evolution of a DOG, we focus on the phase of relatively high total luminosity (i.e., t $\simeq$ 1100 -- 1110 Myr) during which a large amount of gas remains around the AGN.\par

\subsection{Infrared spectrum}\label{sec:sed}
\begin{figure}[ht!]
	\begin{center}
		\includegraphics[width=8cm, bb=0 0 190mm 254mm]{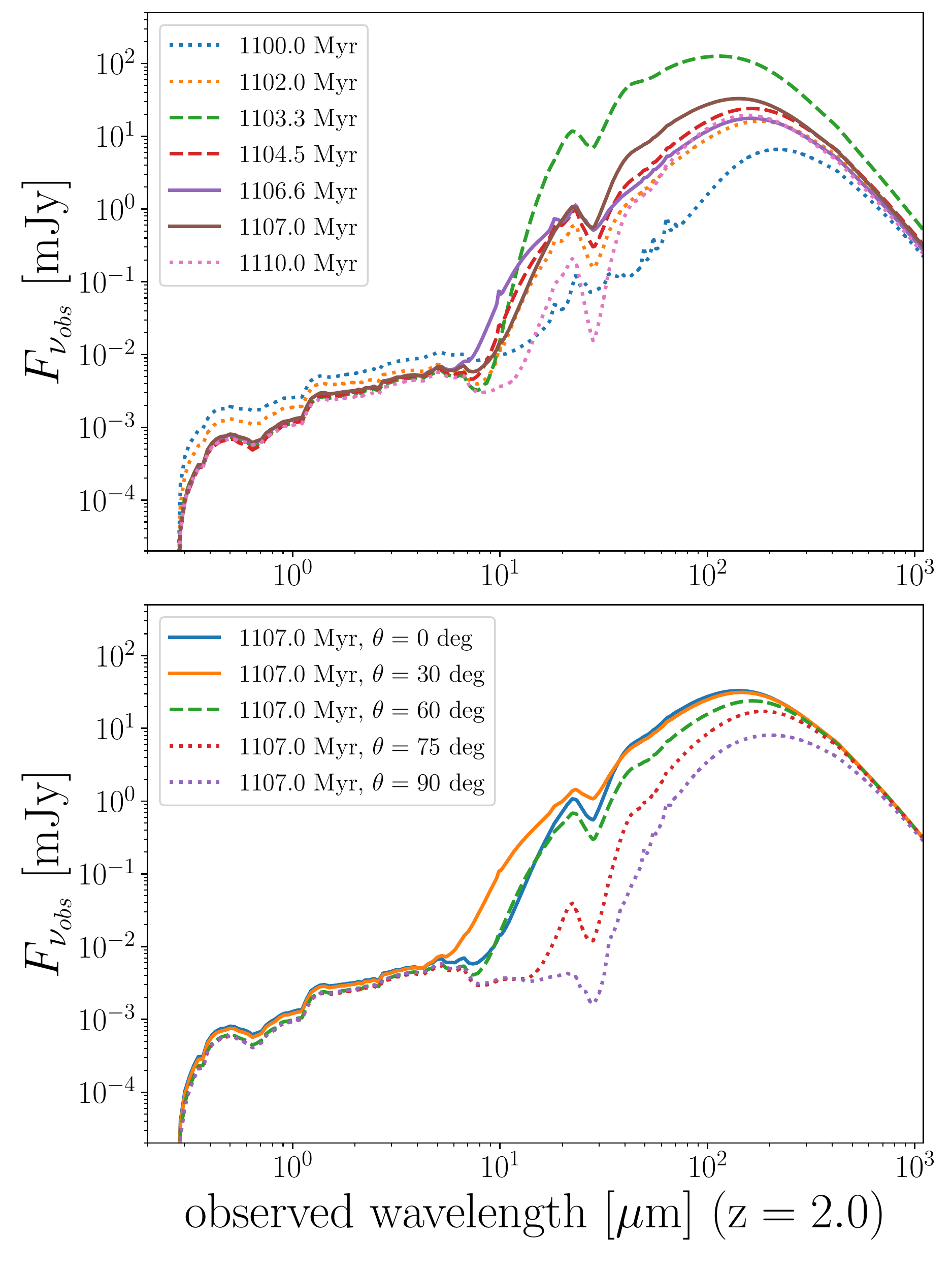}
	\end{center}
	\caption{SEDs obtained using \texttt{RADMC-3D}.
		The redshift of the model was assumed to be z = 2.0.
		Solid, dashed, and dotted lines represent PL DOGs, bump DOGs, and non-DOGs respectively.
		Following \citet{narayanan2010}, the bump DOGs and PL DOGs are classified by the intensities at 3.6, 4.5, and 7.9 $\rm \mu m$: $S_{3.6}\,<\,S_{4.5}$ and $S_{4.5}\,>\,S_{7.9}$ for bump DOGs and $S_{3.6}\,<\,S_{4.5}\,<\,S_{7.9}$ for PL DOGs.
		Top panel : Time evolution of the SED at $\theta$ = 0 deg.
		As the mass accretion rate increases, the SED becomes brighter in the mid-infrared.
		Bottom panel : Comparison of the SEDs observed from $\theta$ = 0, 30, 60, 75, and 90 deg at 1107.0 Myr.
		The SED for $\theta\,>\,60$ deg becomes fainter in the mid-infrared.
		Note that the SEDs in the near-infrared are not sensitive for the viewing angles.
		This is because of the vertical distribution of the stars and gas.
		See Figure \ref{fig:app_image} in Appendix \ref{sec:app_nir}.\label{fig:sed}}
\end{figure}

Figure \ref{fig:sed} shows how the SED of the merger system depends on time and viewing angle $\theta$ for the observer.
We perform pseudo-observations assuming that the model is at z = 2.0, because the population density of DOGs peaks around z $\sim$ 2 \citep{dey2008}.
Here, we distinguish the SEDs obtained at different times and viewing angles using the DOGs criteria (i.e., $F_{24\mu m}/F_{\rm R}$ $\geq$ 982 and $F_{24\rm{\mu m}}\,\geq\,0.3\rm{\,mJy}$).
In addition, as in \citet{narayanan2010}, we classify bump DOGs and PL DOGs by their intensities at 3.6, 4.5, and 7.9 $\rm \mu m$ because bump DOGs have a peak around 4.8 $\rm \mu m$ in the observed frame ($z \sim 2$) (see Figure \ref{fig:sed} caption for details).
From the upper panel of Figure \ref{fig:sed}, we can see that SED changes significantly from 1102.0 Myr to 1103.3 Myr.
Between 1102.0 Myr -- 1103.3 Myr, the system becomes about ten times brighter in the infrared (8 -- 1000 $\mu$m).
This corresponds to the change in the total AGN luminosity from 7.91 $\times$ 10$^{45}$ $\rm erg\,s^{-1}$ to 1.09 $\times$ 10$^{47}$ $\rm erg\,s^{-1}$ during 1.3 Myr (middle panel of Figure \ref{fig:bhdis&macc}).
This rapid change in luminosity can be expected from the dynamical timescales in the merger system.
The middle panel of Figure \ref{fig:bhdis&macc} shows that the total AGN luminosity varies on a timescale of $\sim$ 0.1 Myr.
This corresponds to the dynamical timescale of the nuclear region.
The Kepler time period (i.e., $T=\sqrt{4\pi^2r^3/GM_{BH}}$) at $r\ =$ 4.0 pc from the SMBH with $M_{BH} = 10^8\rm{\ M_{\odot}}$ is about 0.075 Myr.
Infrared luminosity increase was caused by the emission from hot dust near two AGNs.
In other words, the infrared brightness of the merger system depends on the activity of the AGNs.
\citet{toba2017b} also reported a correlation between infrared luminosity and AGN activity for IR Bright DOGs.\par

From the bottom panel of Figure \ref{fig:sed}, we can see that the brightness in the infrared region \textit{decreases} significantly as the viewing angle increases.
This is because the radiation from hot dust in the vicinity of the AGNs is more severely attenuated by the material near the disk plane when the observer is closer to edge-on.
Note that all SEDs shown in this panel were obtained at the same timestamp; therefore, the material distribution and covering factor around the AGNs are the same for the SEDs, and only the viewing angle is changed.
When the viewing angle is larger than 75 deg, the AGNs appear to be faint in the infrared because the dust thermal emission is self-shielded by the foreground cold dust, and consequently the system no longer satisfies the DOGs criteria.
The effect of the viewing angle depends on the wavelength in the infrared.
For example, the dependence in the near-infrared region is smaller than that in the mid-infrared. 
This is because the optical to near IR in the observed frame is dominated by stars, whose scale height is much larger than that of the cold gas/dust that obscures the nucleus (see Appendix \ref{sec:app_nir}).
Figures \ref{fig:image_8.0} compares the maps of the 24 $\rm \mu m$ intensity S$_{24}$ (Jy pixel$^{-1}$) at 1107.0 Myr for $\theta$ = 0 and 75 deg.
At $\theta$ = 0 deg, it was confirmed that the mid-IR emission was dominated by the vicinity of the AGNs (r $<$ 10 pc).
In contrast, at $\theta$ = 75 deg, the radiation from the warm dust in the polar direction was dominant.\par

\begin{figure}[ht!]
	\begin{center}
		\includegraphics[width=16cm, bb=0 0 508mm 381mm]{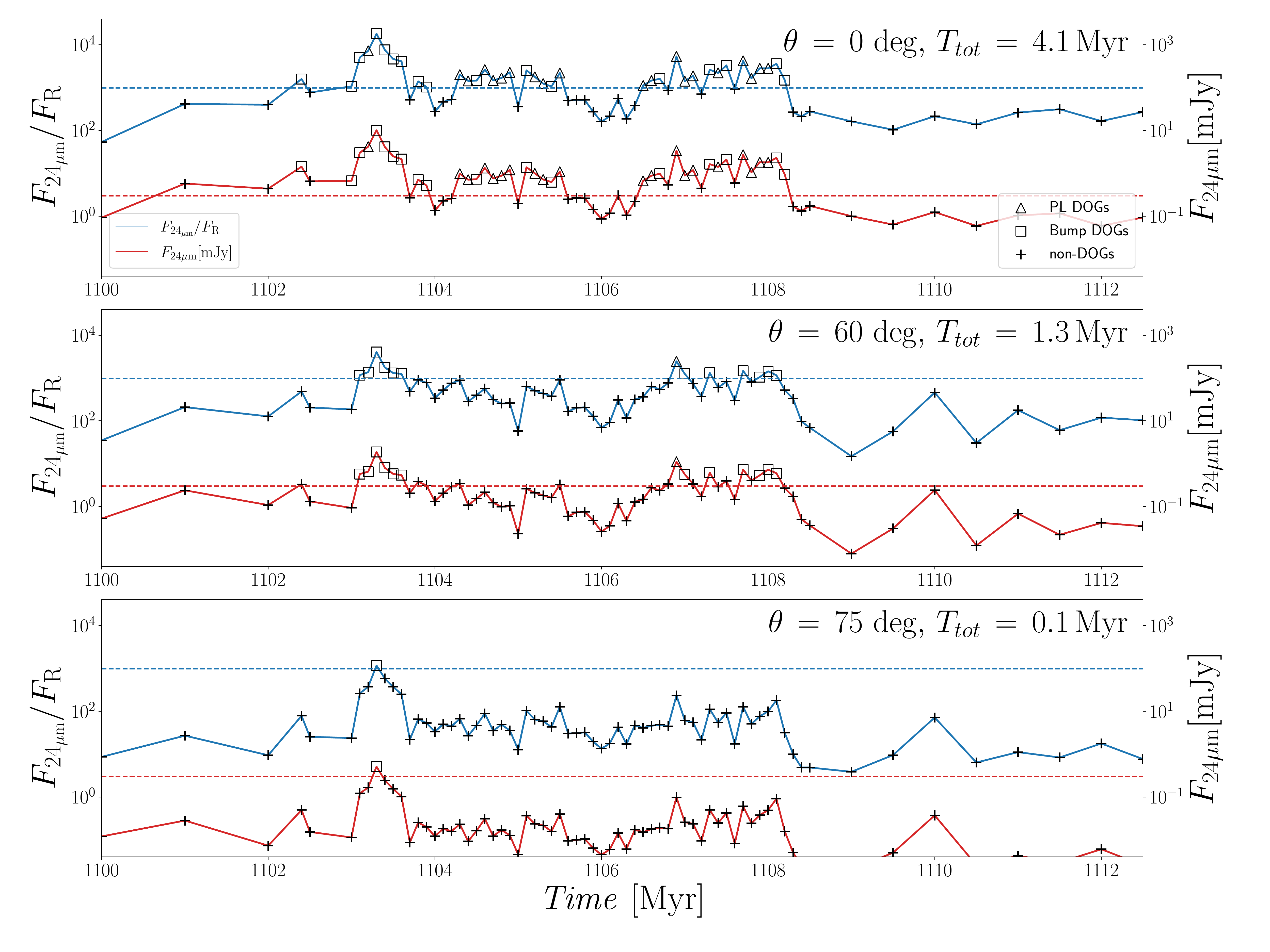}
	\end{center}
	\caption{Comparison of the time evolutions of $F_{\rm 24\mu m}/F_{\rm R}$ and $F_{24\rm{\mu m}}$ in different viewing directions. From top to bottom, $\theta\ =$ 0, 60 and 75 deg.
	The plus signs (+) denote points that do not satisfy the DOGs criteria.
	Squares ($\square$) denote bump DOGs, and triangles ($\triangle$) denote the PL DOGs.
	The redshift of the model was assumed to be z = 2.0.
	The bule and red dotted lines show the DOGs criteria of $F_{\rm 24\mu m}/F_{\rm R}$ = 982 and $F_{24\rm{\mu m}}$ = 0.3 mJy, respectively.\label{fig:lifetime}}
\end{figure}

Figure \ref{fig:lifetime} shows the time evolution of $F_{\rm 24\mu m}/F_{\rm R}$ and $F_{24\rm{\mu m}}$ at different viewing angles.
Hereafter, we use $T_{tot}$ to denote the total duration of all periods in which the system is categorized as a DOGs (i.e., $F_{24\mu m}/F_{\rm R}$ $\geq$ 982, $F_{24\rm{\mu m}}\,\geq\,0.3\rm{\,mJy}$).
When observed face-on ($\theta\ =$ 0 deg), $T_{tot}$ is about 4.1 Myr, but for $\theta\ =$ 60 deg, it is about 1.3 Myr.
Furthermore, from $\theta\ =$ 75 deg, it becomes even shorter, that is, $T_{tot}$ is approximately 0.1 Myr.
This is mainly because $F_{24\rm{\mu m}}$ becomes faint \textit{at large $\theta$}, as shown in Figure \ref{fig:sed}, and fails to meet the DOGs flux threshold.
These results suggest that the lifetime of a DOGs depends on the amount of cold dust ($T_{dust}\ <\ 100$ K) between the system and the observer as well as the time scale of mass accretion, that is, how long hot dust ($T_{dust}\ <\ 300$ K) is sustained by the AGN feedback (see above).
The amount of cold dust in the line of sight and the contribution from the hot dust near AGNs are the key factors that determine whether a system is apparently classified as a DOG.\par

In top panel ($\theta$ = 0 deg) of Figure \ref{fig:lifetime}, it is noteworthly that in the early part of the DOG phase the merger system is observed more frequently as a bump DOG than a PL DOG, while in the late part it becomes vice versa.
In other words, the system is found to evoluves from a bump DOG to a PL DOG.
\Add{Althogh this result similar the evolution in the mid-IR SEDs from bump-like to PL-like discovered by Narayanan et al. 2010, the detail time evolution and transition between the phases were not clear owing to the limited time resolution in their simulations. Our results shows the evolution from bump DOGs to PL DOGs with sufficient time resolution.}
\Add{We have confirmed the transition from bump DOGs to PL DOGs with better time resolution (10 Myr vs. 0.1 Myr).}
In the early DOGs phase, the emission from hot dust ($T_{dust}$ $>$ 300 K) hardly contributed to the flux at 8 $\mu m$ in the observed frame because the AGN was deeply buried (see Figure \ref{fig:gas}), resulting in the dominance of bump DOGs.
However, in the late DOGs phase, PL DOGs are dominant because the gas in the disks is lost due to star formation, thermal AGN feedback, and angular momentum transportation.
\Add{In other words, the evolution from bump DOGs to PL DOGs is affected by AGN-feedback.
We confirmed that if the AGN feedback efficiency is increased from $\eta_{AGN}$ = 0.2\% in the fiducial model to 0.8\%, the lifetime of bump DOGs phase becomes shorter, although the evolution from bump DOGs to PL DOGs in our model does not change (see Appendix \ref{sec:app_eta_agn}).\footnote{\Add{To be note that no bump DOGs phase is observed for $\eta_{AGN}$ = 2.0\%.}}}
\Add{In addition}, for $\theta$ = 60 deg (middle panel of Figure \ref{fig:lifetime}), PL DOGs do not frequently appear even in the late DOGs phase.
In summarly PL DOGs tend to be identified when observed face-on in the late DOGs phase in our model.

\begin{figure}[ht!]
	\begin{center}
		\includegraphics[width=8cm, bb=0 0 190mm 317mm]{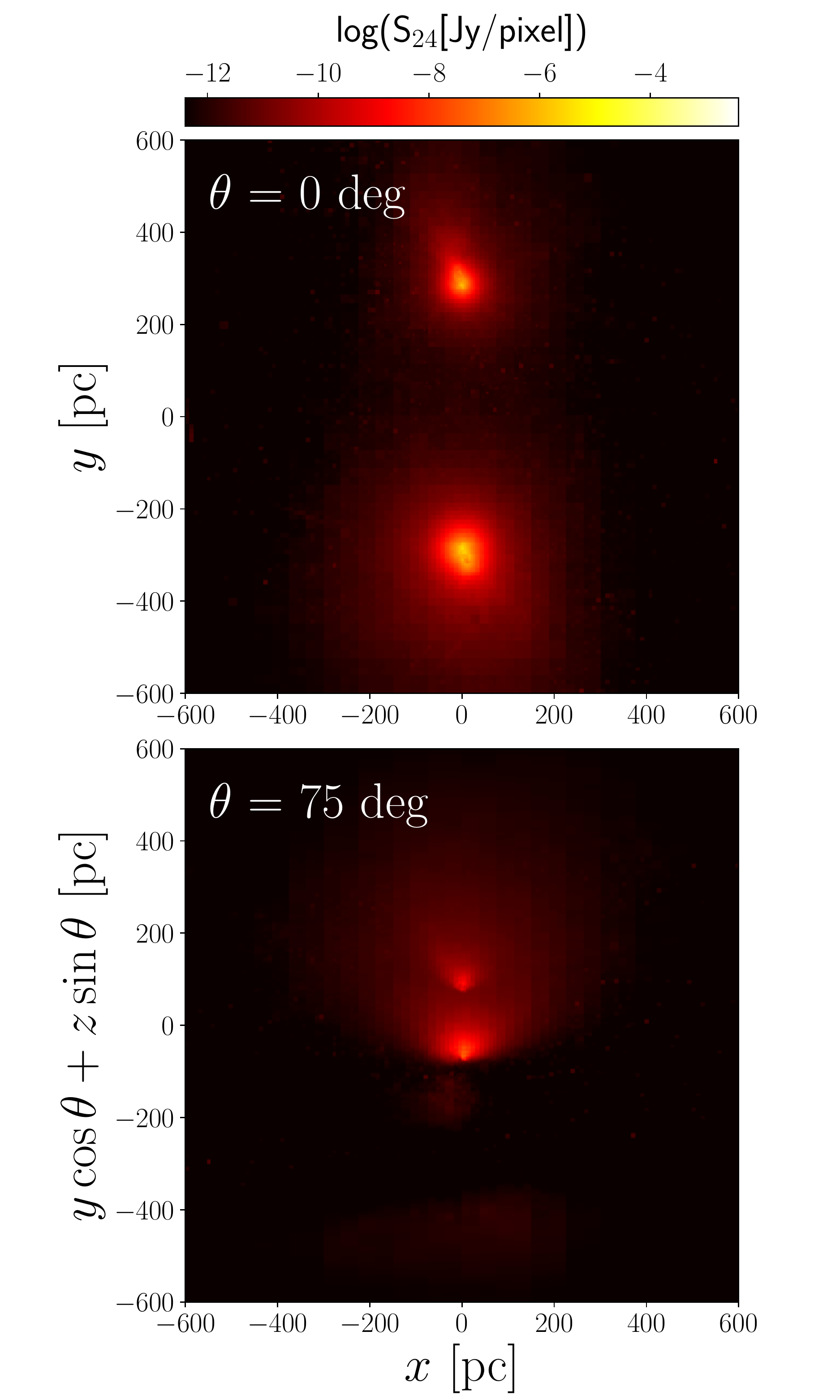}
	\end{center}
	\caption{intensity distribution at 24 $\rm{\mu m}$ in the observed-frame (z = 2.0) at t = 1107.0 Myr. Top panel : $\theta$ = 0 deg, Bottom panel : $\theta$ = 75 deg.\label{fig:image_8.0}
	}
\end{figure}

\subsection{Energy source for Merger-driven DOGs}\label{sec:agn-activity}
\begin{figure}[ht!]
	\begin{center}
		\includegraphics[width=8cm, bb=0 0 163mm 122mm]{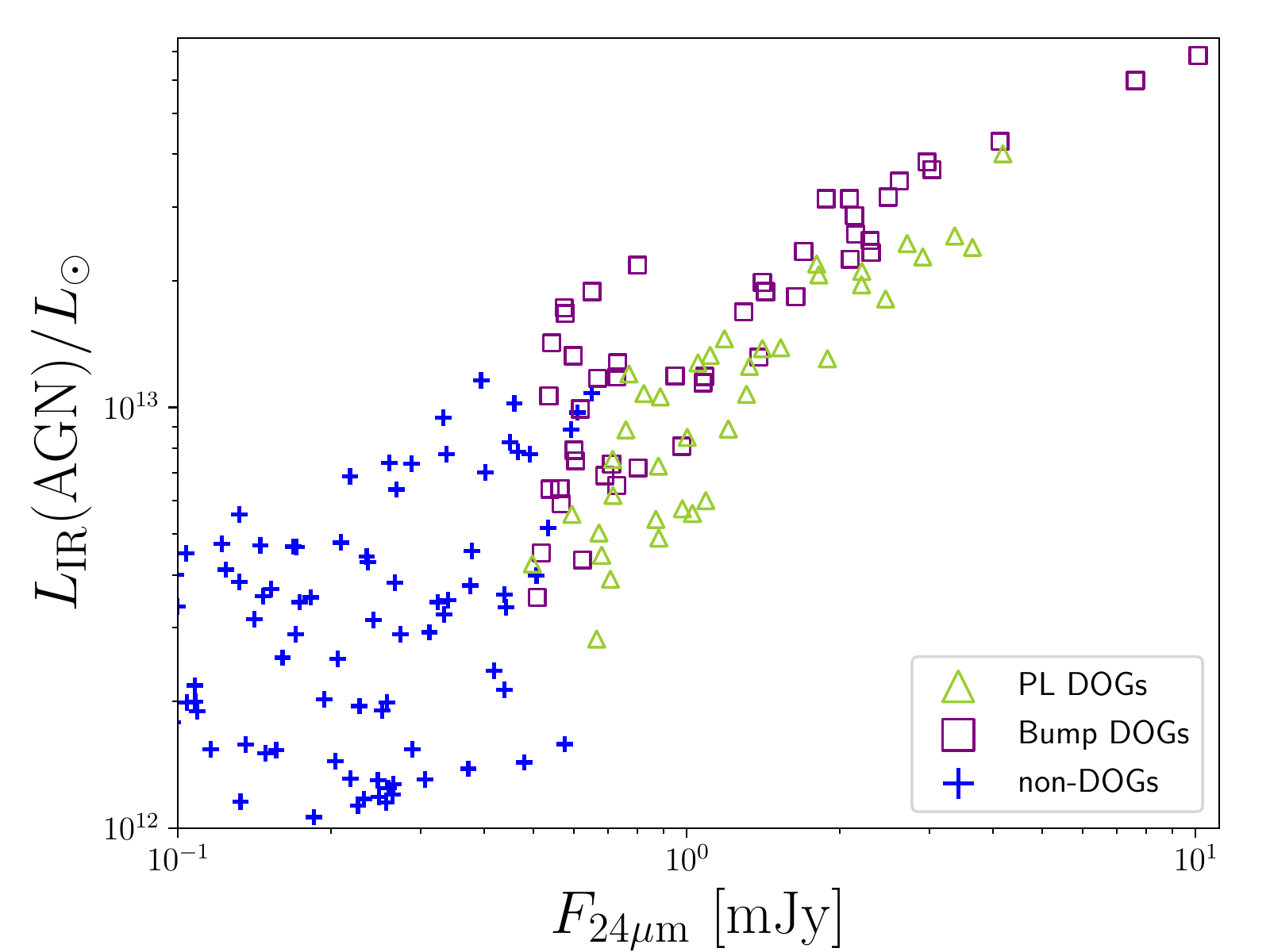}
	\end{center}
	\caption{infrared (8--1000 $\rm \mu m$) luminosity of AGN origin vs $F_{24\rm{\mu m}}$ for $\theta = $ 0, 30, 60, 90 deg during 1100 to 1115 Myr.
	$L_{\rm{IR}}$(AGN) does not include the contributions from stars.
	Plus signs denote timestamps that do not satisfy the DOGs criteria.
	Squares denote bump DOGs, and triangles denote the PL DOGs.
	$L_{\rm{IR}}$(AGN) is positively correlated with $F_{24\rm{\mu m}}$.
	\label{fig:f24}}
\end{figure}

PL DOGs have often been considered AGN-dominated sources in the infrared.
One indirect piece of evidence for this interpretation is a positive correlation between the PL DOGs population ratio and $F_{24\rm{\mu m}}$ \citep{dey2008,melbourne2012,toba2015}.
To confirm the origin of PL and bump DOGs, we plotted the infrared luminosity arising from AGNs $L_{\rm IR}$(AGN) versus $F_{24\rm{\mu m}}$ in Figure \ref{fig:f24}.
$L_{\rm IR}$(AGN) is calculated by running \texttt{RADMC-3D} without stellar radiation and integrating the resultant spectrum over 8--1000 $\rm \mu m$.
Clearly $F_{24\rm{\mu m}}$ positively correlates with $L_{\rm IR}$(AGN), implying that the mid-infrared brightness represents the contribution from the AGNs in DOGs.
The correlation coefficients for PL and bump DOGs were approximately 0.927 and 0.921, respectively.
Thus, in our model, both PL and bump DOGs are strongly affected by AGN luminosity in the mid-IR.
However, at the same AGN luminosity, $F_{24\rm{\mu m}}$ of PL DOGs tends to be brighter than that of the bump DOGs.
The differences between PL and bump DOGs are discussed in Section \ref{sec:dis_column}.\par

\section{ Discussion } \label{sec:discuss}
\subsection{Evolution of Merger-driven DOGs}\label{sec:dis_column}
In Section \ref{sec:results}, we show that the merger system in our model has the same observational properties as DOGs, albeit depending on the viewing angle.
Contrary to intuition, the merger system is more likely to be observed as DOGs based on their color and IR flux if it is closer to \textit{face-on}.
This is because if the system is close to edge-on, the hot dust around the AGNs which is responsible for producing a sufficiently high IR flux, is self-shielded by the foreground cold and dense gas, which is concentrated near the disk plane.
As described in Section \ref{sec:sed}, the top panel of Figure \ref{fig:image_8.0} shows the intensity distribution of $I_{24\rm{\mu m}}$ in the observed frame when observed face-on ($\theta = $0 deg) at 1107.0 Myr.
The top panel of Figure \ref{fig:image_8.0} also shows that a bright core in the mid-infrared region is associated with each SMBH.
It should be noted that the current observations cannot resolve the structures of DOGs at $z > 1$ on a sub-kpc scale.
Therefore, we currently have no observational information on the spatial structure of the DOGs' central region.
We discuss this based on the results of our numerical simulation.\par

\begin{figure}[ht!]
	\begin{center}
		\includegraphics[width=8cm, bb=0 0 19.05cm 31.75cm]{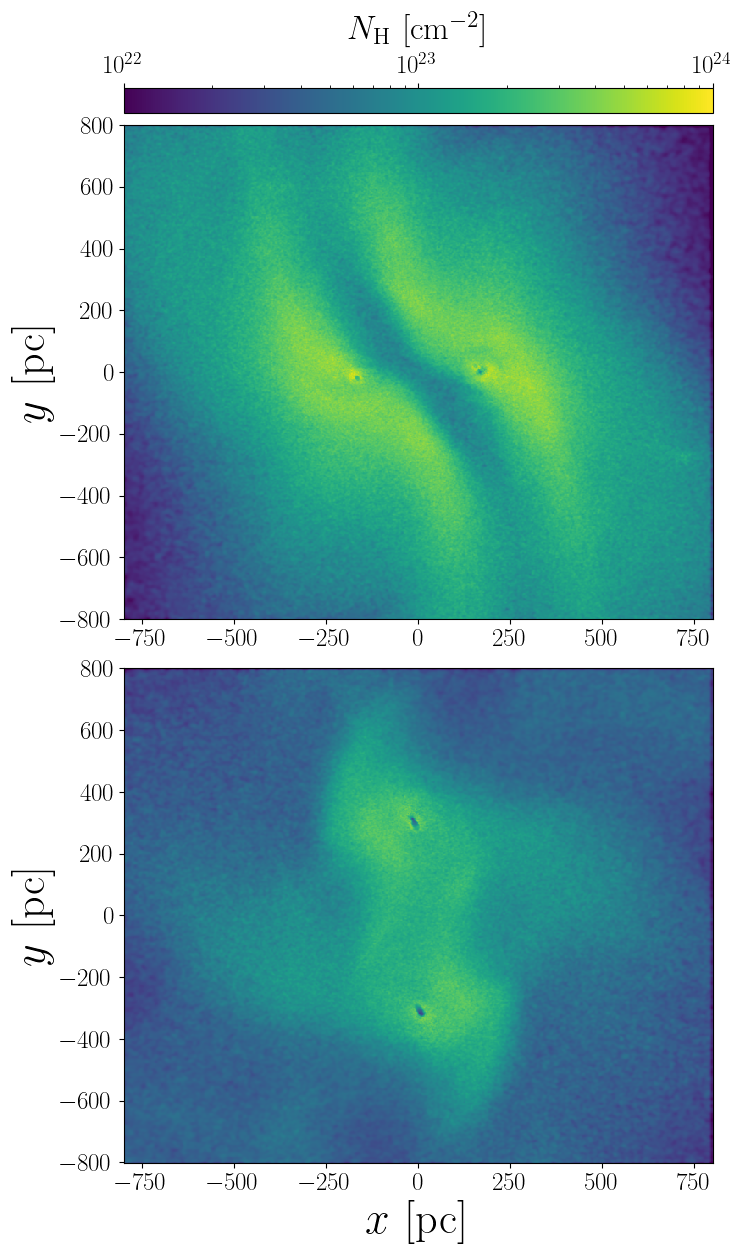}
	\end{center}
    \caption{Map of column density projected to the x-y plane at t = 1103.3 Myr (top panel) and 1107.0 Myr (bottom panel) \label{fig:NH_2D}}
\end{figure}

Figure \ref{fig:NH_2D} shows the gas column density projected onto the x-y plane at t = 1103.3  Myr and 1107.0 Myr. 
At t = 1103.3 Myr, we confirmed that the AGNs were buried in gas with at column density of 5 $\times$ 10$^{23}$ cm$^{-2}$ or greater.
At that time, a cavity of 10 pc was formed by feedback from each AGN.
However, at t = 1107.0 Myr, the column density averaged at r $<$ 100 pc is lower than 5 $\times$ 10$^{23}$ cm$^{-2}$ and AGN feedback cavities is seen more prominently.\par

\begin{figure}[ht!]
	\begin{center}
		\includegraphics[width=8cm, bb=0 0 19.05cm 31.75cm]{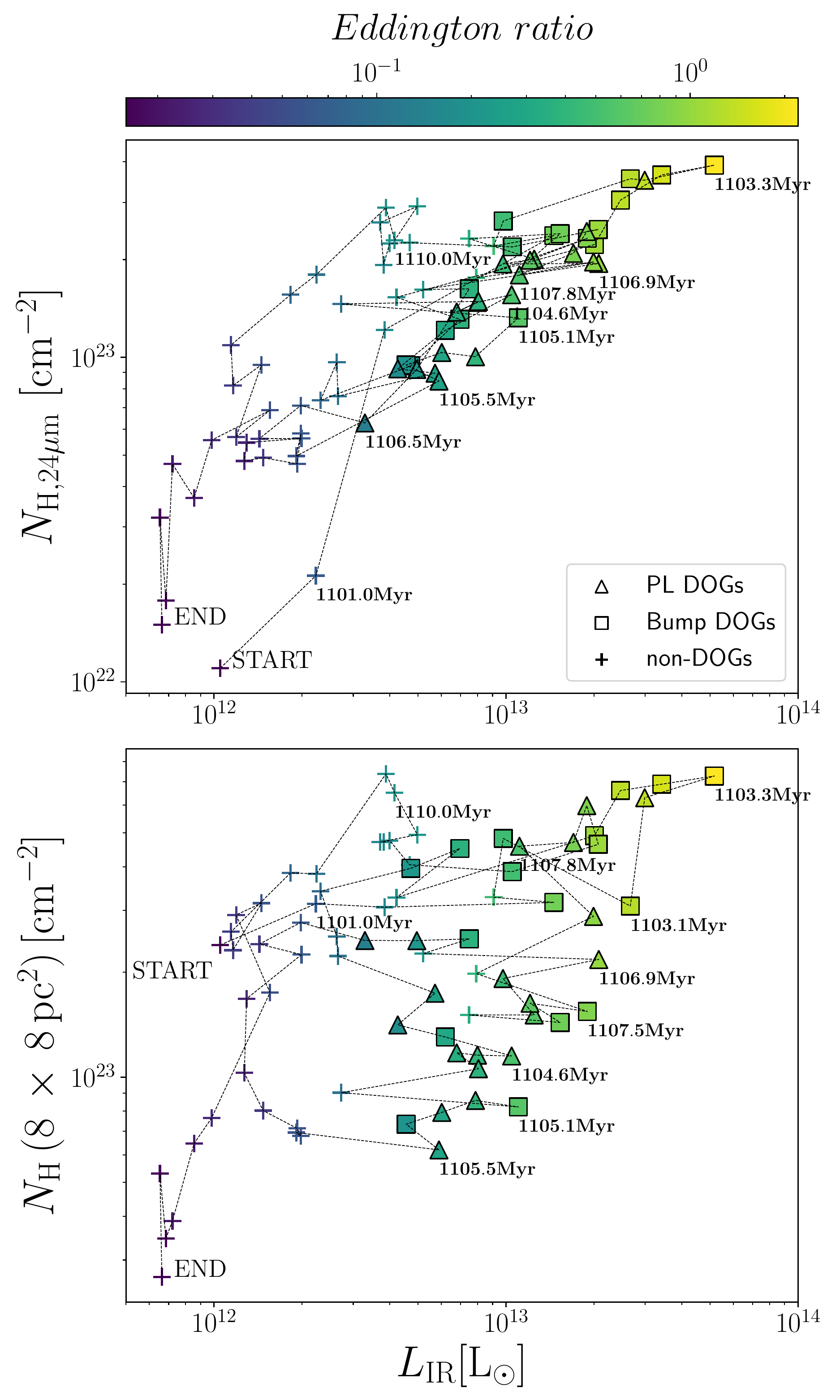}
	\end{center}
	\caption{Top panel : $N_{\rm H, 24\rm \mu m}$ vs $L_{\rm{IR}}$(8 -- 1000 $\rm{\mu m}$),
		Bottom panel : $N_{\rm H} {\rm (8 \times 8\ pc^2)}$ vs $L_{\rm{IR}}$(8 -- 1000 $\rm{\mu m})$
		We assumed z = 2.0, and the plots were observed from $\theta$ = 0 deg.
		The color bar represents the Eddington ratio.
		The plus signs do not satisfy the DOGs criteria.
		Squares indicate bump DOGs and triangles indicate PL DOGs.\label{fig:NH_LIR}}
\end{figure}

The top panel of Figure \ref{fig:NH_LIR} shows the time evolution of the column density when observed face-on.
Here, the 24 $\rm \mu m$ intensity weignted mean column density $N_{\rm H,24\mu m}$ is defined as
\begin{equation}\label{eq:NH}
	N_{\rm H,24\mu m} = \frac{\int\int N_{\rm H}(x,y)\ S_{\rm 24\mu m}(x,y)dx dy}{\int\int S_{\rm 24\mu m}(x,y)dxdy},
\end{equation}
We obtain the $S_{\rm 24\mu m}$ [Jy/pixel] distribution using \texttt{RADMC-3D} (e.g., Figure \ref{fig:image_8.0}).
The top panel of Figure \ref{fig:NH_LIR} shows that the column density varied by an order of magnitude between 1101.0 Myr and 1103.3 Myr.
During this period, the AGNs became buried with the supplied gas, and the Eddington ratio also increased by more than one order of magnitude.
From 1103 Myr to 1104 Myr, AGN is deeply buried and the system is observed as a bump DOG (see Figure \ref{fig:NH_LIR}).
Subsequently, the gas near the SMBH is blown out by AGN feedback, and $N_{\rm H,24\mu m}$ and L$_{\rm IR}$ decrease.
These results refine the bump-to-PL DOGs evolution inferred from Figure \ref{fig:lifetime} in Section \ref{sec:sed}.
As a result, the system was observed as PL DOGs from 1104 Myr to 1108 Myr.\par

We also confirmed the evolution of the central column densities in the 8pc from AGN.
The bottom panel of Figure \ref{fig:NH_LIR} shows the evolution of the central column density, which is evaluated in the 8 $\times$ 8 $\rm{pc^2}$ box centered on SMBH, and the infrared luminosity.
The column densities were averaged for the two SMBHs.
We note that the simulated N$_{\rm H}$ and Eddington ratio well explains the distribution of eROSITA-detected obscured AGNs \citep{toba2021}.
We can also confirm that there are two populations: PL DOGs with $N_{\rm H}$ $\sim$ 10$^{23}$ cm$^{-2}$ and, bump DOGs with $N_{\rm H}$ $\sim$ 5 $\times$ 10$^{23}$ cm$^{-2}$.
This difference occur because the gas around the nucleus is consumed by the star formation and mass accretion to the nucleus.\par

\subsection{Mid-infrared color selection for Merger-driven DOGs}\label{sec:twocolor}
Most luminous infrared galaxies are in the phase of advanced mergers and are powered by a mixture of circumnuclear starburst and AGNs \citep{sanders1996}. 
However, it is not straightforward to quantify the contribution of starburst and AGNs to their SEDs observationally.
For example, in our case, both the star formation rate and mass accretion rate increased when the two galaxies approached each other (see Figure \ref{fig:bhdis&macc}).\par

It is known that mid-infrared color selection is helpful for identifying AGNs \citep{jarrett2011,mateos2012}.
This is because the dust in AGNs can be heated up to the sublimation temperature ($\leq$ 1500 K) by the intense UV radiation from the AGN.
Thermal re-emission from the dust causes a power-law spectrum in the infrared range (3 -- 30 $\rm \mu$m); therefore, the difference in their colors can be used to infer the contribution of the AGNs.\par

\citet{blecha2018} proposed a new color selection scheme of AGN using WISE for nearby ULIRGs.
They also showed the evolution of nearby ULIRGs on the WISE two-color diagram.
However, DOGs are generally observed at high redshifts (z $>$ 1) \citep{dey2008}.
The classification on the WISE two-color diagram ([3.4] -- [4.6] $>$ 0.5, [4.6] -- [12] $>$ 2.2, and [3.4] -- [4.6] $>$ 2.0 $\times$ [4.6] -- [12] - 8.9) of the nearby ULIRGs (z $<$ 1) cannot be directly applied to DOGs.\par

\begin{figure}[ht!]
	\begin{center}
		\includegraphics[width=8cm, bb=0 0 190mm 317mm]{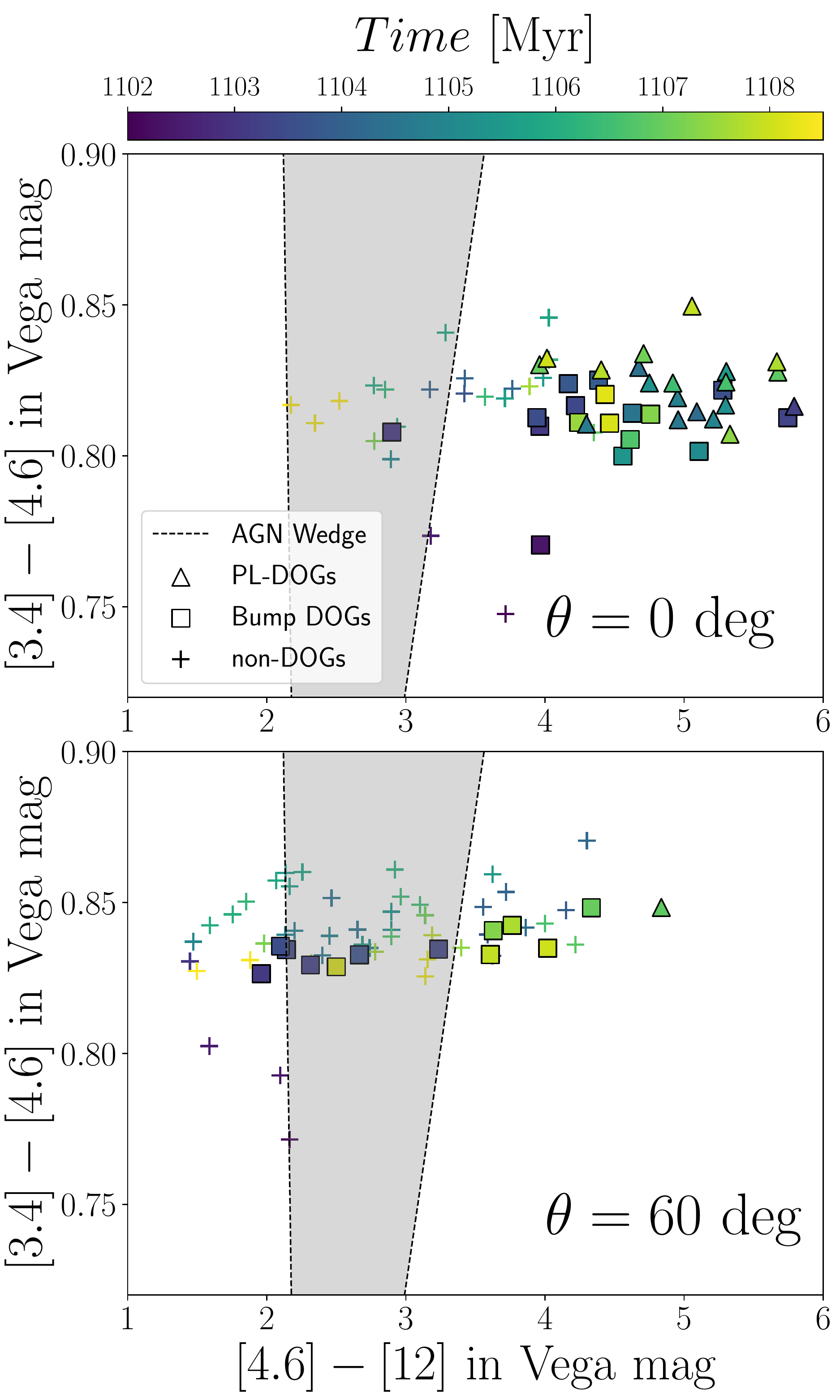}
	\end{center}
	\caption{Time evolution of the system on the two-color diagram (z = 2.0). 
		The AGN Wedge is represented in gray \citep{mateos2012}.
		The plus signs do not satisfy the DOGs criteria. 
		The meanings of the plus, square, and triangle symbols are the same as Figure \ref{fig:NH_LIR}.
		Top panel: $\theta\ =$ 0 deg, bottom panel: $\theta\ =$ 60 deg\label{fig:twocolor}}
\end{figure}

In Figure \ref{fig:twocolor}, we plot evolutional tracks of DOGs (z = 2.0) on the WISE two-color diagram.
This shows that DOGs are distributed both inside and outside the AGN-Wedge \citep{mateos2012}.
Therefore, this results suggest that AGN-Wedge on the WISE two-color diagram does not necessarily correspond to merger-driven DOGs.
The mid-infrared color varies significantly with $\theta$, and PL DOGs tend to be redder than bump DOGs in [4.6] -- [12].
In addition, for $\theta$ = 60 deg, [4.6] -- [12] tends to be bluer than for $\theta$ = 0 deg. 
[4.6] -- [12] is sensitive to radiation from the AGN, whereas [3.4] -- [4.6] show no significant change during the DOGs phase.
Our results indicate that DOGs color selection is challenging.\par

\section{summary and conclusions}
\Add{We simulated the final phase of galaxy collisions resolved dust torus scale (several tens pc) using the N-body/SPH code \texttt{ASURA} \citep{saitoh2008, saitoh2009, saitoh2013} to explore the origin of DOGs.}
Snapshots obtained from the N-body/SPH simulation were used as inputs to the radiation transport simulation code \texttt{RADMC-3D} \citep{dullemond2012} to investigate the evolution of the SED including its dependence on the viewing angle.
The goal of this simulation is to determine the power source of merger-driven DOGs and to clarify the relationship between PL DOGs and bump DOGs.
The main results are as follows:

\begin{enumerate}
	\item In the final phase of galaxy mergers, we have found snapshots that satisfy the spectral characteristics of bump DOGs and power-law DOGs. (see Sections \ref{sec:sed} and \ref{sec:twocolor})
	\item The SED changes significantly during a few Myrs correspoding to the timescale of mass accretion to r $<$ 4 pc. (see Section \ref{sec:sed})
	\item The lifetime of DOGs changes depending on the viewing angles because the contribution of AGN is attenuated by high column density of dusty gas as the viewing angle increases (i.e., closer to edge-on). (see Section \ref{sec:sed})
	\item For merger driven DOGs, we have confirmed that bump DOGs evoluve to PL DOGs. This result is consistent with the senario proposed by \cite{dey2009}. \Add{\cite{narayanan2010} also reported a similar transition in the mid-IR SEDs. We have confirmed the transition of DOGs type with better time resolution of 0.1 Myr. Furthermore, the transition was observed even when AGN-feedback efficiency $\eta_{AGN}$ was increased from 0.2\% to 0.8\%. We note that as $\eta_{AGN}$ is increased, the lifetime of bump DOGs in the early DOGs phase will be shorter, and the bump DOGs phase is no longer observed for $\eta_{AGN}$ = 2.0\%. (see Section \ref{sec:sed} \& Appendix \ref{sec:app_eta_agn}) }
\end{enumerate}

As a final remark, one should note that the numerical treatment of triggering AGNs during the gas-rich merger of our simulations is still idealized, in the sense that we do not crrectly solve detailed physical processes within 4 pc of the SMBH. 
The current model of the fueling process to the AGNs and the feedback from the AGNs are simplified.
We are planning to perform 3D hydrodynamic simulations inside the 4 pc near the SMBH using the mass accretion rate in this study as a boundary condition to resolve the structure within 4 pc from SMBH in the late stages of galactic collision.
We believe that this will allow us to better assess the structure of AGNs in the late stages of galactic collisions.

\begin{acknowledgments}
We would like to thank Takayuki Saitoh for providing \texttt{ASURA} code.
This work was supported by the Japan Society for the Promotion of Science (JSPS) KAKENHI Grant Numbers 16H03959 and 21H04496.
Y. T. was supported by JSPS KAKENHI Grant Numbers 18J01050, and 19K14759.
Numerical computations were performed on a Cray XC50 at the Center for Computational Astrophysics, National Astronomical Observatory of Japan.
\end{acknowledgments}

%

\software{ASURA \citep{saitoh2008,saitoh2009,saitoh2013},  
	      RADMC-3D \citep{dullemond2012}, 
	      Cloudy \citep{ferland1998}, 
	      P\'egase.3 \citep{fioc2019}}



\appendix

\section{ Near infrared images and SEDs }\label{sec:app_nir}
\begin{figure}
	\begin{center}
		\includegraphics[width=18cm, bb=0 0 279mm 76mm]{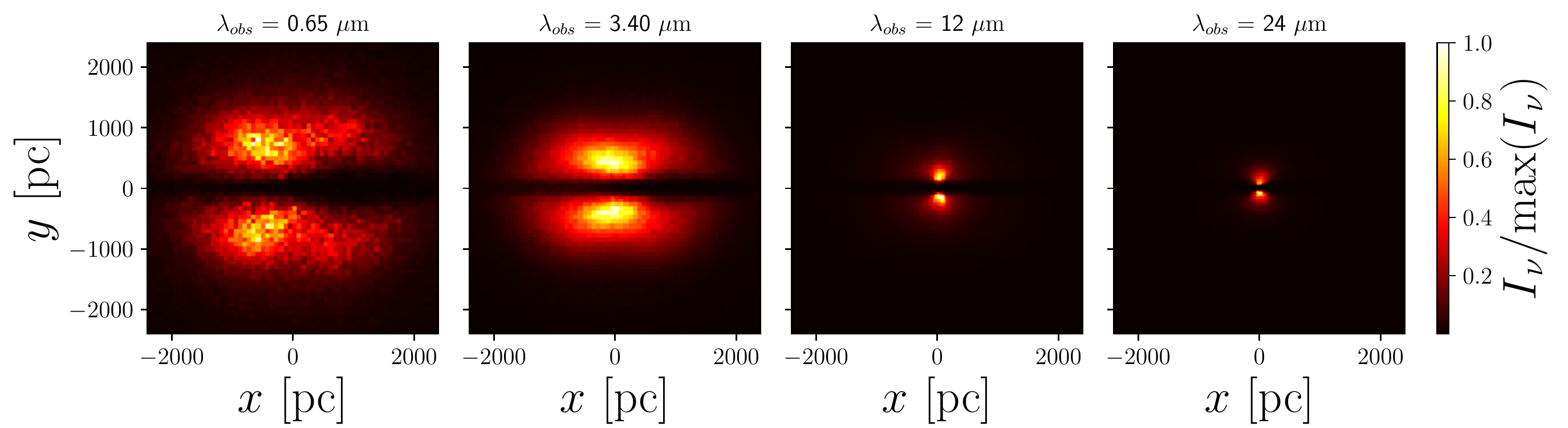}
	\end{center}
	\caption{Intensity distribution in optical to mid-infrared for $\theta$ = 90 deg at t = 1107.0 Myr.\label{fig:app_image}}
\end{figure}

The SEDs shown in Figure \ref{fig:sed} (bottom) do not exhibit significant changes in the near-IR at t = 1107.0 Myr.
This is because radiation from the stellar system is dominant in the near-infrared, and AGNs are buried in the foreground cold dust and gas.
Figure \ref{fig:app_image} shows that the shorter the wavelength, the more important the radiation from the high-scale-height regions.

\section{ Higher AGN-feedback efficiency }\label{sec:app_eta_agn}
\begin{figure}
	\begin{center}
		\includegraphics[width=16cm, bb=0 0 508mm 254mm]{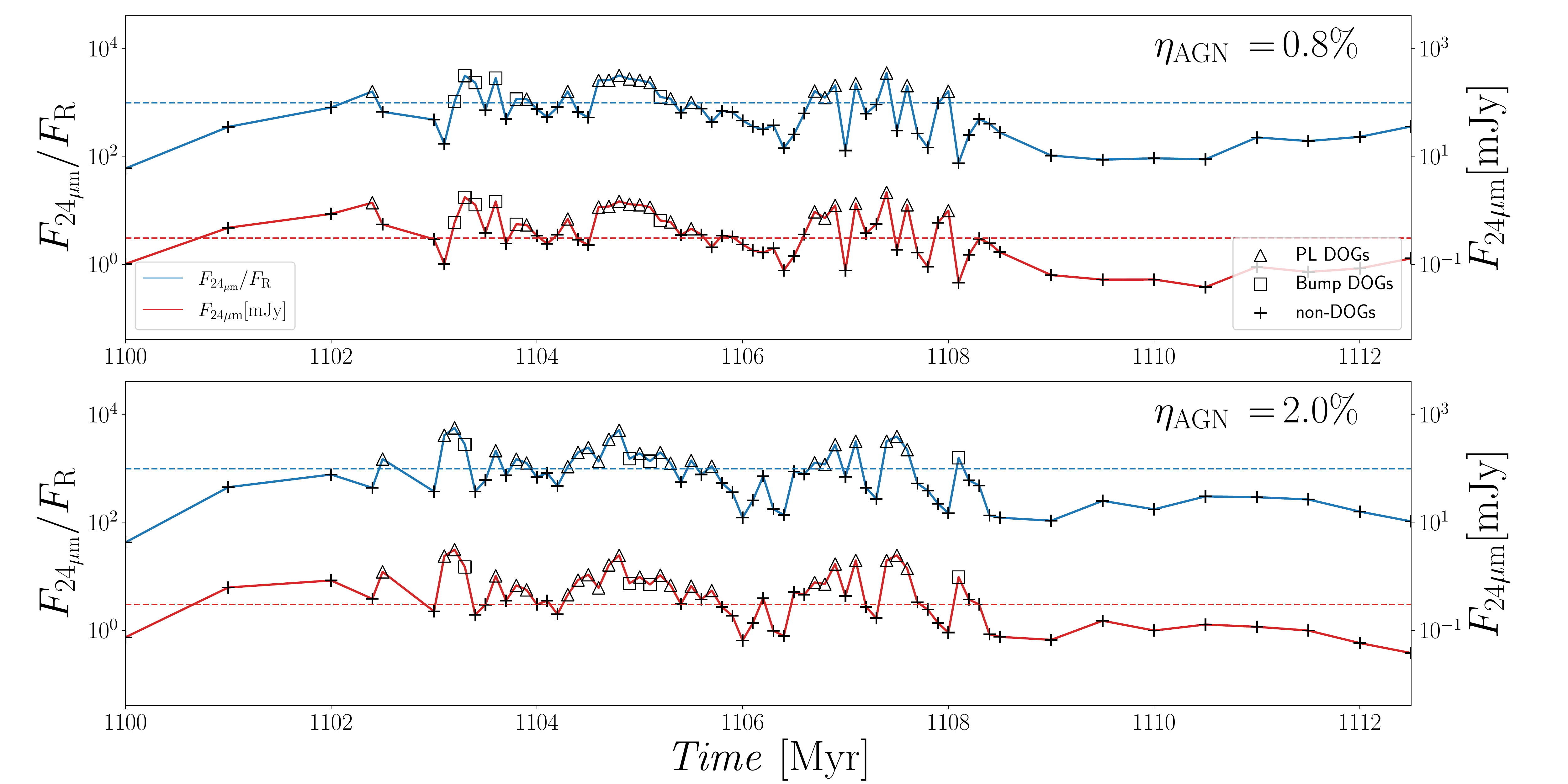}
	\end{center}
	\caption{\Add{DOGs type evolution for $\theta = 0$ deg with varying AGN-feedback efficiency ($\eta_{AGN}$). In the top pane, $\eta_{AGN}$ is 0.8$\%$; in the bottom panel,  $\eta_{AGN}$ is 2.0$\%$.}\label{fig:app_agn}}
\end{figure}

\Add{We have confirmed how higher AGN-feedback efficiency ($\eta_{AGN}$) affects the evolution of DOGs. Figure \ref{fig:app_agn} compare the time evolution of $F_{\rm 24\mu m}/F_{\rm R}$ and $F_{24\rm{\mu m}}$ between $\eta_{AGN}$ = 0.8\% and 2.0\%.  The evolution from bump DOGs to PL DOGs was confirmed for the AGN feedback efficiency of 0.8\% as well as for the fiducial model ($\eta_{AGN}$ = 0.2\%). However, we note that the lifetime of bump DOGs for $\eta_{AGN}$ = 0.8\% is shorter than that of the fiducial model (see Figure \ref{fig:lifetime}), and no bump DOGs phase is observed for $\eta_{AGN}$ = 2.0\%. In other words, as AGN-feedback efficiency increases, PL DOGs are observed more frequently in the early DOGs phase. This is probably because the more powerful AGN feedback blows away the cold dust on the line of sight, and thereby it becomes brighter in the near-infrared.}


\bibliography{yutani}{}
\bibliographystyle{aasjournal}


\end{document}